\pdfoutput=1

\documentclass[11pt,a4paper]{scrartcl}

\usepackage{CLICdp}

\title{Sensitivity of CLIC at 380 GeV to the top FCNC decay
                 \boldmath $t\rightarrow cH$}

\clicdpconf{2017}{005}  

\date{\formatdate{13}{3}{2017}}

\addauthor{A.F.~\.Zarnecki\thanks{Filip.Zarnecki@fuw.edu.pl}}{\institute{1}}

\addinstitute{1}{Faculty of Physics, University of Warsaw, Poland}

\onbehalfof{the CLICdp collaboration}

\abstract{
In the Standard Model (SM), flavour changing neutral current (FCNC) top
decays, possible at loop level only, are very strongly suppressed.
Observation of any such decay would be a direct signature of physics
beyond the SM. Large enhancements are possible in many "new physics"
scenarios and the largest enhancement is in most cases expected for
the $t\rightarrow cH$ decay. A full study for CLIC was based on the
WHIZARD simulation of FCNC top decays within the 2HDM(III) model. Beam
polarization and beam-induced background were taken into account. Top
pair production events with the FCNC decay $t\rightarrow cH$ can be
identified based on kinematic constrains and flavour tagging
information. Due to a large overlap in the kinematic space with
standard top pair events, the final signal selection-efficiency is
small, at the 10\% level. Expected limits on 
$BR(t\rightarrow cH)\times BR(H\rightarrow b\bar{b})$ are compared
with earlier results based on parton level simulation.  
}

\titlecomment{Talk presented at
  the International Workshop on Future Linear Colliders (LCWS2016),\\
  Morioka, Japan, 5-9 December 2016.\\
          C16-12-05.4.}

\graphicspath{ {./logos/}{./plots/} }

\begin{document}


\titlepage

%
%

\section{Introduction}

Top physics, together with Higgs boson studies and searches for Beyond
the Standard Model (BSM) phenomena, is one of the three pillars of the
research program for future high energy $e^{+} e^{-}$ colliders.
As the top quark is the heaviest known elementary particle, with an
expected value of the Yukawa coupling of the order of one, the precise
determination of its properties is a key to the understanding of
electroweak symmetry breaking. 
Determination of top properties is also essential for many ``new
physics'' searches, as the top quark gives large loop contributions to many
precision measurements sensitive to BSM effects.
Stringent constraints on the ``new physics'' scenarios are also
expected from direct searches for rare top decays. 
Both future linear colliders, the International Linear Collider (ILC)
and the Compact Linear Collider (CLIC), provide an opportunity to
study the top quark with unprecedented precision via direct production
of $t\bar{t}$ pairs in  $e^{+} e^{-}$ collisions.
Presented in this contribution are prospects of constraining
the branching ratio for the flavour changing top decay
$t \rightarrow c H$ with CLIC running at $\sqrt{s} = 380$ GeV.

\section{Experimental conditions}

The Conceptual Design Report (CDR) for CLIC
was presented in 2012 \cite{Aicheler:2012bya}.
CLIC is based on a two-beam acceleration scheme, required to
generate a high RF gradient of about 100 MV/m. 
In the recently updated implementation plan for CLIC \cite{CLIC:2016zwp},
a construction in three stages is proposed, with 5 to 7 years of data
taking at each stage.
The first stage with a footprint of 11 km will allow one to reach a
center-of-mass energy of 380 GeV, giving access to most Higgs boson
and top quark measurements. 
The plan assumes collecting 500~fb$^{-1}$ at 380~GeV with additional
100 fb$^{-1}$ collected at the $t\bar{t}$ threshold. 
The CLIC baseline design includes polarisation for the electron beam, while
positron polarisation is considered as a possible upgrade.

The detector concepts proposed for CLIC are based on 
jet reconstruction and jet energy measurements with the ``Particle
Flow'' approach \cite{Thomson:2009rp}. 
Single particle reconstruction and identification exploits high
calorimeter granularity, and the best possible jet energy estimate is
obtained by combining calorimetric measurements of neutral particles
with precise track momentum measurements for the charged ones.
Very efficient flavour tagging is possible with a high precision pixel vertex
detector placed very close to the interaction point.
The background to processes with missing energy can be strongly suppressed
thanks to very good detector hermeticity, with instrumentation extending
down to a minimum angle of $\theta_{\mathrm{min}} \sim 10$~mrad.
Presented in this contribution are detailed simulation results based
on the ILD detector concept adopted for CLIC~\cite{Linssen:2012hp}. 

\section{Theoretical expectations}

\subsection{Standard Model}

Flavour-Changing Neutral Current (FCNC) top quark decays,
$t \rightarrow  c X$ ($X = \gamma, \; g, \; Z, \; H$),
are strongly suppressed in the Standard Model.
Only charged current decays are allowed at the tree level
and the loop level contributions are suppressed by the
Glashow-Iliopoulos-Maiani (GIM) mechanism~\cite{Glashow:1970gm}. 
The cancellation is not perfect because of the non-zero
down-quark masses and the dominant contribution to the FCNC decays
comes from diagrams including a $b$ quark in the loop. 
The corresponding partial widths are proportional to the
square of the element $V_{cb}$ of the Cabibbo-Kobayashi-Maskawa (CKM)
quark-mixing matrix~\cite{Cabibbo:1963yz,Kobayashi:1973fv}
and to the fourth power of the $b$ quark to the $W$ boson mass ratio.
These two suppression factors result in extremely small 
branching ratios expected~\cite{Agashe:2013hma} for the Standard Model:
\begin{eqnarray*}
BR(t  \rightarrow  c g ) &  \sim & 5 \cdot 10^{-12},  \\
BR(t  \rightarrow  c \gamma ) &  \sim & 5 \cdot 10^{-14}, \\
BR(t  \rightarrow  c Z ) &  \sim & 1 \cdot 10^{-14},  \\
BR(t  \rightarrow  c H ) &  \sim & 3 \cdot 10^{-15}. 
\end{eqnarray*}
Observation of any such decay would be a direct signature
for ``new physics''.

One should note that for one of the diagrams contributing to the
$t \rightarrow  c H$ channel the GIM mechanism is not strictly applicable
due to the Higgs coupling being proportional to the quark mass.
But the contribution of this diagram is suppressed by the $b$ quark to
$W$ boson mass ratio, resulting from the Higgs coupling. 
So, in spite of the GIM mechanism violation, the expected FCNC 
branching ratio for this channel is smallest in the Standard Model.

\subsection{Beyond the Standard Model}

Many extensions of the Standard Model predict significant enhancement
of the FCNC top decays~\cite{Agashe:2013hma}.
The enhancement can be due to the direct tree level FCNC coupling, but
in most models it is observed at the loop level and results from
contributions of new particles or from the modified particle couplings.
The decay $t \rightarrow  c H$ seems to be the most promising
channel, as most BSM scenarios predict significant deviation in the
(light) Higgs boson couplings or contributions from the additional
Higgs bosons to the loop diagrams.  
For the Two Higgs Doublet Model (2HDM), which is one of the simplest
extensions of the Standard Model, loop contributions can be 
enhanced up to the level of $BR \sim   10^{-4}$~\cite{Bejar:2001sj}.
For the ``non standard'' scenarios, 2HDM(III) or ``Top~2HDM'',
where one of the Higgs doublets couples to the top quark only,  tree level
FCNC couplings are also allowed and an enhancement up to
$10^{-2}$ is possible~\cite{DiazCruz:2006qy}. 
The  $t \rightarrow  c H$  decay is also difficult to constrain
at the LHC, because of the high QCD background to the dominating Higgs
boson decay channels. The expected limit from HL-LHC is
$BR(t  \rightarrow  c H ) < 2 \cdot 10^{-4}$
\cite{Agashe:2013hma,ATL-PHYS-PUB-2016-019}. 

\subsection{Parton level study}

Before a detailed analysis based on the full detector simulation
results could be started, a feasibility study was performed at the
parton level, with very coarse modeling of detector effects.
Signal and background event samples were generated with
WHIZARD~\cite{Kilian:2007gr,Moretti:2001zz} using an internal CIRCE1
option for modeling the beam energy spectra.
Signal events were generated using the dedicated implementation of
2HDM(III) in SARAH~\cite{Staub:2015kfa}.
Model parameters were tuned to obtain
$BR(t  \rightarrow  c H ) = 10^{-3}$ for the assumed Higgs boson
mass $m_{H} = 125$~GeV.
The analysis was based on the 6-fermion final state generated by WHIZARD,
with only the $H\rightarrow b\bar{b}$ decay channel considered for the
signal.
As a $t\bar{t}$ sample can be selected at $e^+e^-$ colliders with high
purity, it was also assumed that the dominant background to FCNC events
comes from the standard top quark decay channels.

Top pair production events with the FCNC decay $t \rightarrow  c H$
can be identified based on the kinematic constrains and flavour
tagging information. 
To model the detector performance, a very simplified parametrisation
was used:
\begin{itemize}
\item detector acceptance for leptons: $| \cos \theta_l | < 0.995$;
\item detector acceptance for jets: $| \cos \theta_j | < 0.975$;
\item jet energy smearing: $\sigma_E / E = S/\sqrt{E [\textrm{GeV}]}$ (or
  $S/10$ for jet energy $E>100$ GeV);
\item no energy smearing for reconstructed final state leptons
  (electrons and muons);
\item fixed $b$ quark tagging efficiency $\varepsilon_b$, $c$ and light quark
  misstagging efficiencies,  $\varepsilon_c$ and  $\varepsilon_{uds}$.
\end{itemize}
For the jet energy resolution, three scenarios were considered:
$S=30\%$ (optimistic scenario), $50\%$ (realistic scenario) and $80\%$
(pessimistic scenario).
For the tagging efficiencies, two scenarios, based on LCFI+~\cite{Suehara:2015ura}
results, were eventually used: 
$\varepsilon_b = 80\%$, $\varepsilon_c = 8\%$ and  $\varepsilon_{uds}
= 0.8 \%$ (looser $b$ quark selection) or  $\varepsilon_b = 70\%$,
$\varepsilon_c = 2\%$ and  $\varepsilon_{uds} = 0.2 \%$ (tighter selection). 

\begin{figure}
\begin{center}
\includegraphics[width=0.6\textwidth]{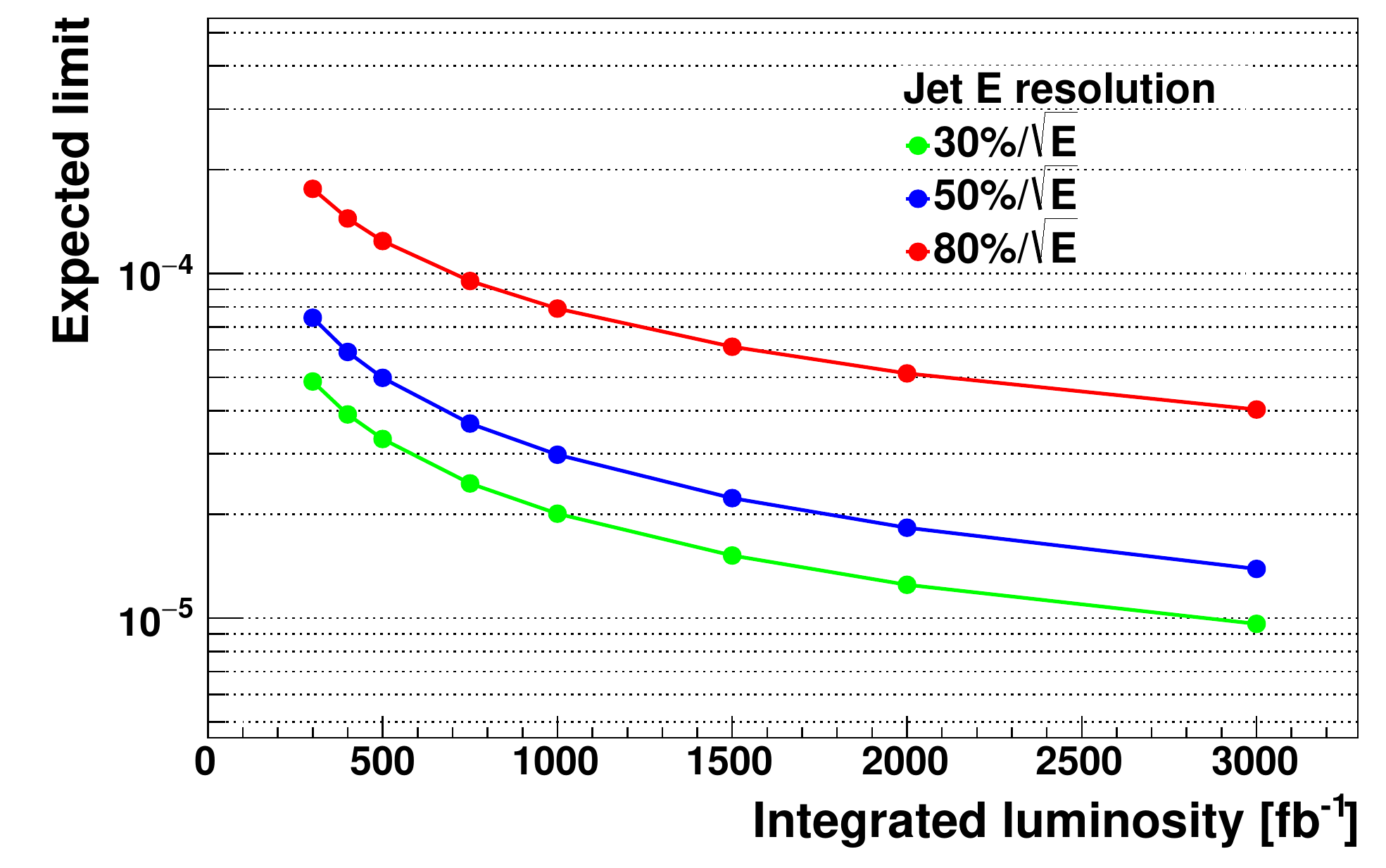}
\end{center}
\caption{\label{plot_res380}Expected 95\% C.L. limits on the
$BR(t  \rightarrow  c H ) \times BR(h  \rightarrow  b  \bar{b})$
  as a function of the integrated luminosity for $e^+e^-$ collisions
  at $\sqrt{s} = 380$~GeV. Results are based on a parton level study
  assuming different jet energy resolution parameters, as indicated in
  the plot. Both hadronic and semi-leptonic decay channels are considered.} 
\end{figure}

\begin{figure}
\begin{center}
\includegraphics[width=0.6\textwidth]{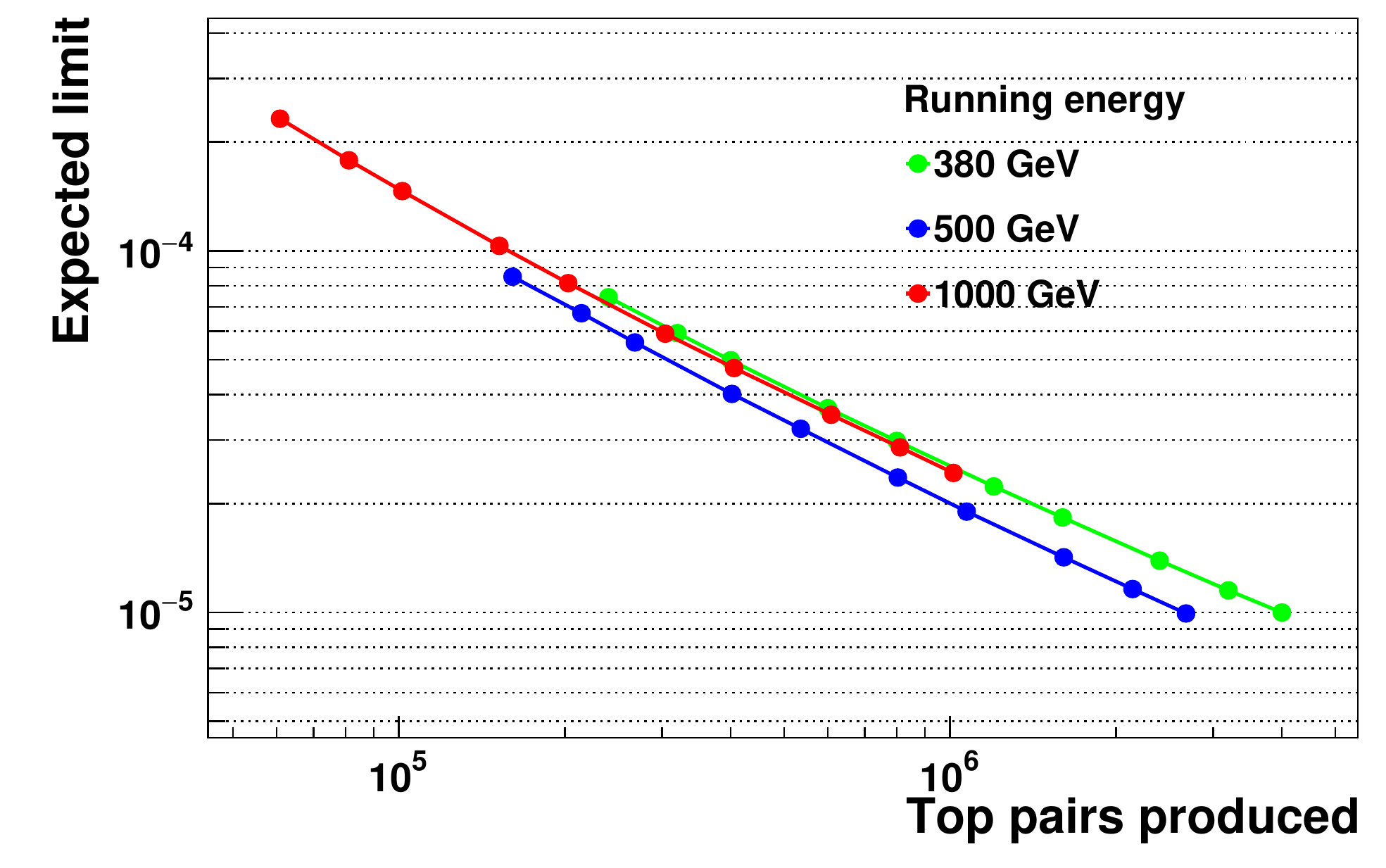}
\end{center}
\caption{\label{plot_num50}Expected 95\% C.L. limits on the
$BR(t  \rightarrow  c H ) \times BR(h  \rightarrow  b  \bar{b})$
  as a function of the collected sample of top pair production events.
  Results are based on a parton level study  for different running
  energies, assuming a jet energy resolution parameter $S = 50\%$.
  Both hadronic and semi-leptonic decay channels are considered.} 
\end{figure}

Both fully-hadronic and semi-leptonic decay channels of the produced top
pairs were considered. After imposing the $b$-tagging criteria (three
tagged jets required: two $b$ from the Higgs boson decay and one from
the ``spectator'' top quark, the one with charged current decay),
a kinematic fit was performed for each event, for both signal and
background hypothesis.   
The $\chi^2$ formula for both cases included constraints on the masses
of two top candidates, $W$ mass constraint and the Higgs boson or
second $W$ mass constraint.
The final selection was based on the ratio of the $\chi^2$ values for
best signal and best background hypothesis.
The selection was optimised so as to obtain the highest sensitivity to
the FCNC decays.
For jet energy resolution given by $S=30\%$ or $50\%$, loose
$b$-tagging settings were chosen while for the worst resolution
($S=80\%$), tighter flavour selection is required.
Expected limits on $BR(t  \rightarrow  c H ) \times BR(h 
\rightarrow  b \bar{b})$, for $e^+e^-$ collisions at $\sqrt{s} =
380$~GeV, as a function of the integrated luminosity are shown in
figure~\ref{plot_res380}. 
Although the signal selection efficiency is small, limited by a large
overlap in the kinematic space with standard top pair events, 
results indicate that with high integrated luminosity this decay can
be probed down to $BR \sim 10^{-5}$~\cite{Vos:2016til}.  
The sensitivity of the linear collider experiments to the top quark FCNC decays
is primarily determined by the available statistics of the 
top quark pairs.
Except for the top pair production cross section, the running energy has only
a marginal influence on the expected branching ratio limit, as illustrated in 
figure~\ref{plot_num50}. 

\section{Event simulation and reconstruction}

Detailed detector level analysis was performed for  $e^+e^-$ collisions
at CLIC, for $\sqrt{s} = 380$~GeV.
As for the parton level study, signal event samples were generated
using the 2HDM(III) model implemented in SARAH.
However, to improve the description of the energy distribution, 
a dedicated CLIC beam spectra file was used.
An electron beam polarisation of 80\% was also taken into account.
Events generated with WHIZARD 2.2.8 were passed to PYTHIA~6.4 for
hadronisation with quark masses and other settings adjusted to the
configuration used previously in CLIC CDR studies~\cite{Linssen:2012hp}.
The generated signal sample was then processed with a standard event
simulation and reconstruction chain of the CLICdp collaboration using
the CLIC\_ILD\_CDR500 detector configuration. 
The background sample considered in the analysis included a full set of
six-fermion event samples produced previously for CLICdp studies of top
pair production at $\sqrt{s} = 380$~GeV.
All sub-samples corresponded to an integrated luminosity of at least
500~fb$^{-1}$. 
The procedure used for signal event generation was also used to
generate an additional ``test'' sample of Standard Model top pair-production
events, which was then used to validate the consistency of simulation
settings between old (background) and new (signal and test) samples.

In the final step, a dedicated event analysis procedure was applied to
all signal and background events.
Because of the large event sample considered,
iLCDirac~\cite{ilcDiracLC} was used as an interface to grid resources,
and the processing was based on the MARLIN~\cite{Gaede:2006pj}
framework with the ilcsoft version v01-17-09.
The reconstructed object collection resulting from loose background
rejection cuts (Loose\-Selected\-Pandora\-PFA\-New\-PFOs) was used as an input
for jet reconstruction with the Valencia
algorithm~\cite{Boronat:2016tgd} as well as for primary and secondary
vertex (re)reconstruction, and flavour tagging with LCFI+.

\section{Selection of signal events}

\begin{figure}
\begin{center}
  \includegraphics[width=0.45\textwidth]{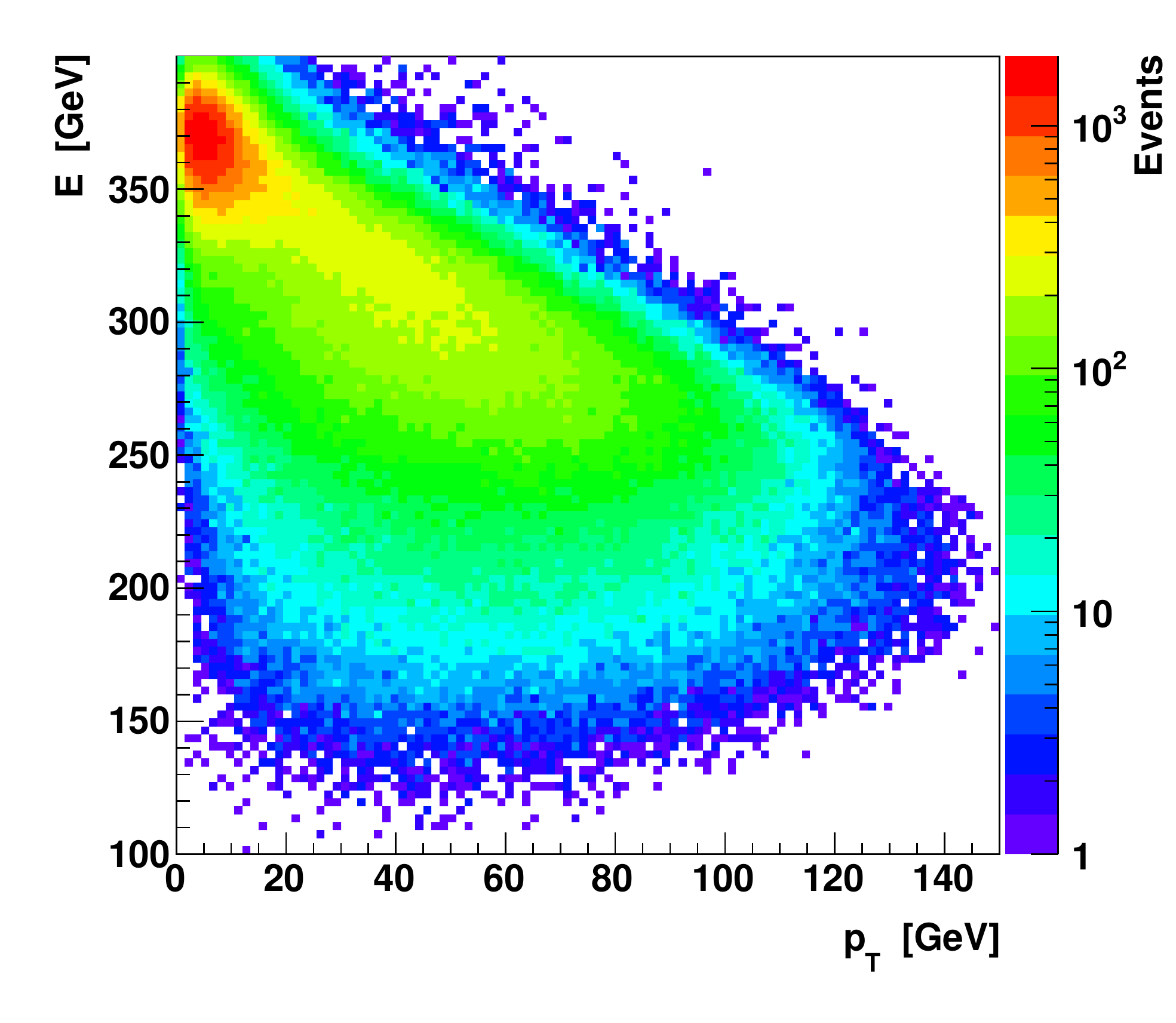}
  \hspace{0.05\textwidth}
\includegraphics[width=0.45\textwidth]{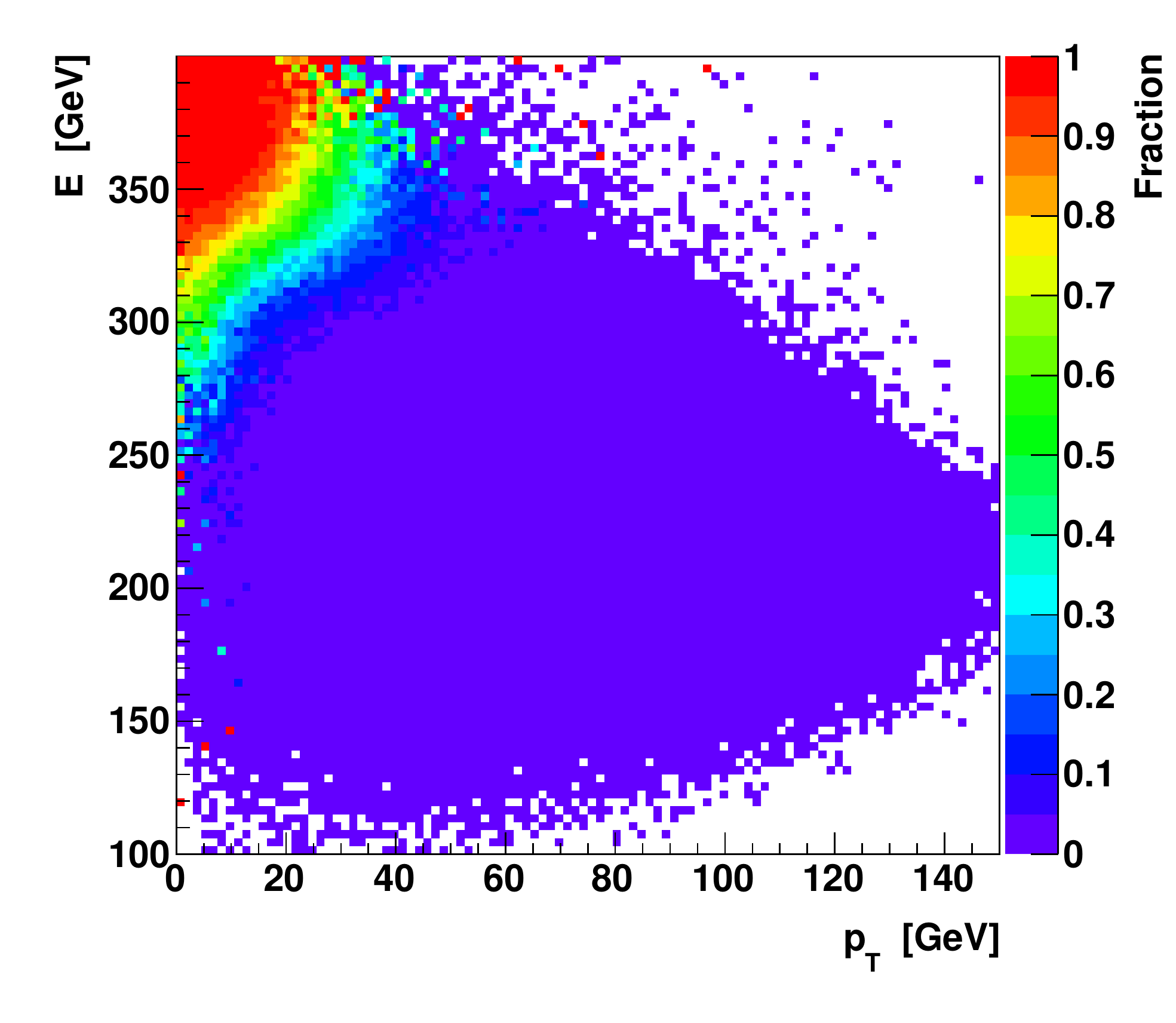}
\end{center}
\caption{\label{had_sel} Left: correlation between the
measured transverse momentum, $p_T$, and the total energy of the
event, $E$, for the background events at $\sqrt{s} = 380$~GeV.
Right: fraction of hadronic events as a function of the  $p_T$ and  $E$. } 
\end{figure}

Results presented in this contribution were based on the analysis of
hadronic top decays only, i.e. events with all $W$ bosons  decaying to
two quarks (six jet final state). 
Hadronic events can be selected by looking at the correlation of the
measured transverse momentum, $p_T$, and the total energy of the
event, $E$, as shown for the considered
background sample in figure~\ref{had_sel} (left). 
Shown in figure~\ref{had_sel} (right) is the probability that the
considered $p_T$ and $E$ values resulted from the hadronic top pair
decay.
It shows that an efficient selection of hadronic events in this plane
can be made with a single cut on the $E - 2 p_T$ value.
The selection efficiency can be still improved slightly by considering
also the total longitudinal momentum of an event, $p_z$.
We defined the effective variable describing the energy balance in the
event as

\[ E_{\mathrm{balance}} = \sqrt{ (E - 2  p_T - \sqrt{s})^2  +  4 p_Z^2}\; , \]
and for hadronic events, $E_{\mathrm{balance}} < 100$ GeV was required.

Before the kinematic fit was used to test the signal vs. background
hypothesis, additional preselection cuts were applied to select 
candidate events.
We required that at least three out of six jets reconstructed with the 
Valencia algorithm were tagged as $b$-quark jets (with tag probability of
at least 0.4) and the fourth jet was tagged as a $c$ or $b$ quark 
(corresponding to $c$ quark from $t\rightarrow cH \rightarrow c b \bar{b}$ decay).
The estimated efficiency of the described preselection cuts is about
34\% for signal events and about 2.4\% for background events
(including the hadronic branching fraction).

The best signal hypothesis was then selected by comparing $\chi^2$
values for different jet assignments.
As significant correlations were observed between the reconstructed Higgs
boson mass and the ``signal'' top quark mass (the one decaying to
$cH$), and between the reconstructed $W$ boson mass and the ``spectator'' top quark
mass, these measurements should not be treated as independent.
Therefore the $\chi^2$ function is defined based on the measured
values of the following parameters:
\begin{itemize}
\item ``signal'' and ``spectator'' top quark masses;
\item ``signal'' and ``spectator'' top quark boosts (energy to mass ratios);
  \item the ratio of the Higgs boson mass to the ``signal'' top quark
    mass;
  \item the ratio of the $W$ boson mass to the ``spectator'' top quark mass.
\end{itemize}
A similar approach was used to select the best background hypothesis
(replacing the Higgs boson by another $W$ and requiring a $b$-tag only
for two jets).

After the kinematic fit, additional selection criteria were imposed to
optimise signal event selection:
\begin{itemize}
\item signal hypothesis fit resulting in $\chi^2_{\mathrm{sig}} < 14$;
\item difference of two reconstructed top masses $\Delta M_t < 45$ GeV;
\item product of $b$-tag probabilities for two jets from Higgs decay
  $b_1 \cdot b_2 > 0.95$;
\item $b$-tag probability for the $b$ jet from spectator top decay
  $b_3 > 0.9$;
\item sum of $c$-tag and $b$-tag probabilities for $c$ quark from
  Higgs decay $c_4 + b_4 > 0.75$.
\end{itemize}
A final cut was then applied on the ratio of $\chi^2$ values for the best
signal and background hypothesis, $\chi^2_{\mathrm{sig}}/\chi^2_{\mathrm{bg}}$.

\section{Results}

\begin{figure}
\begin{center}
\includegraphics[width=0.45\textwidth]{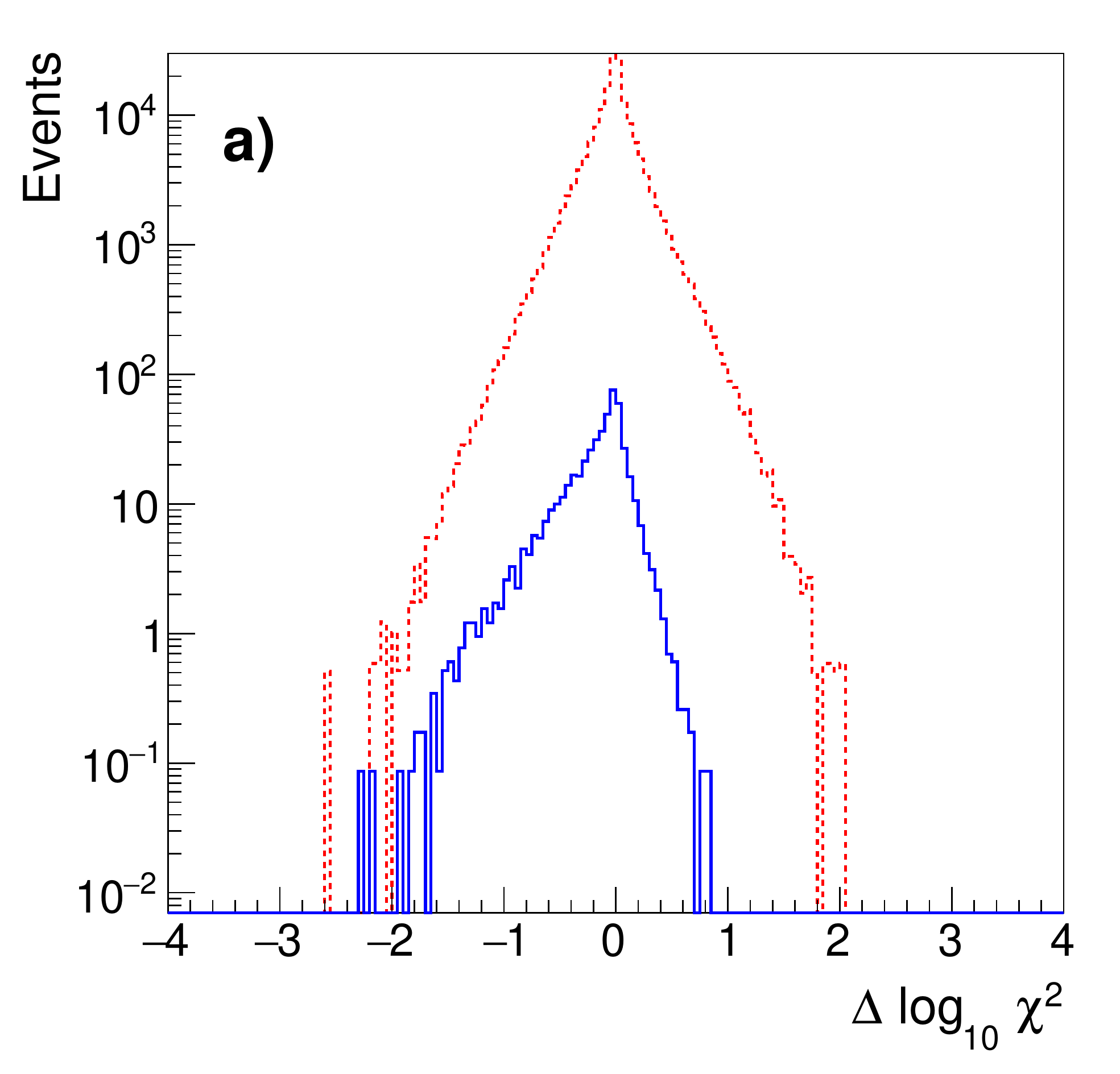}
\includegraphics[width=0.45\textwidth]{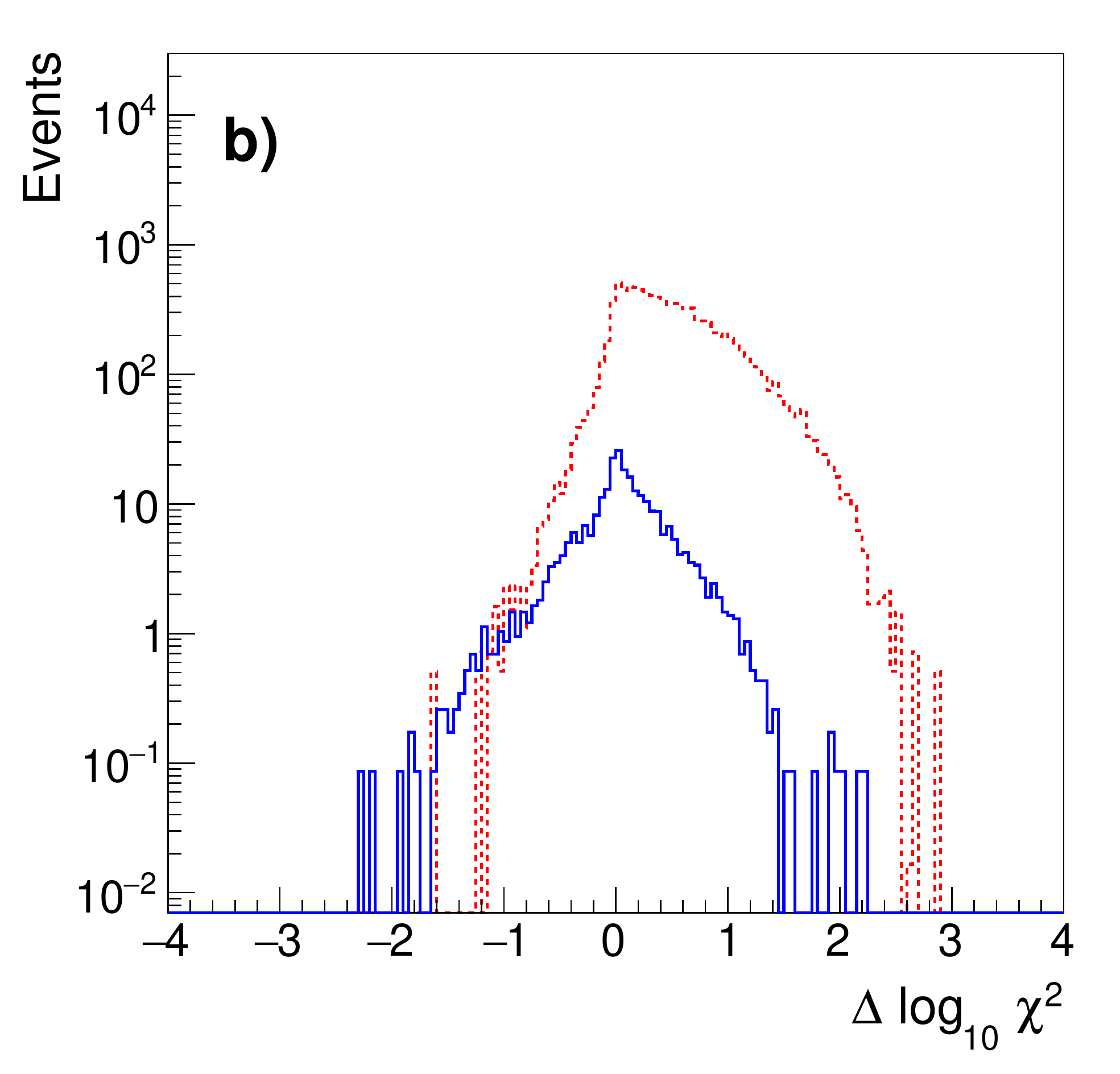}
\includegraphics[width=0.45\textwidth]{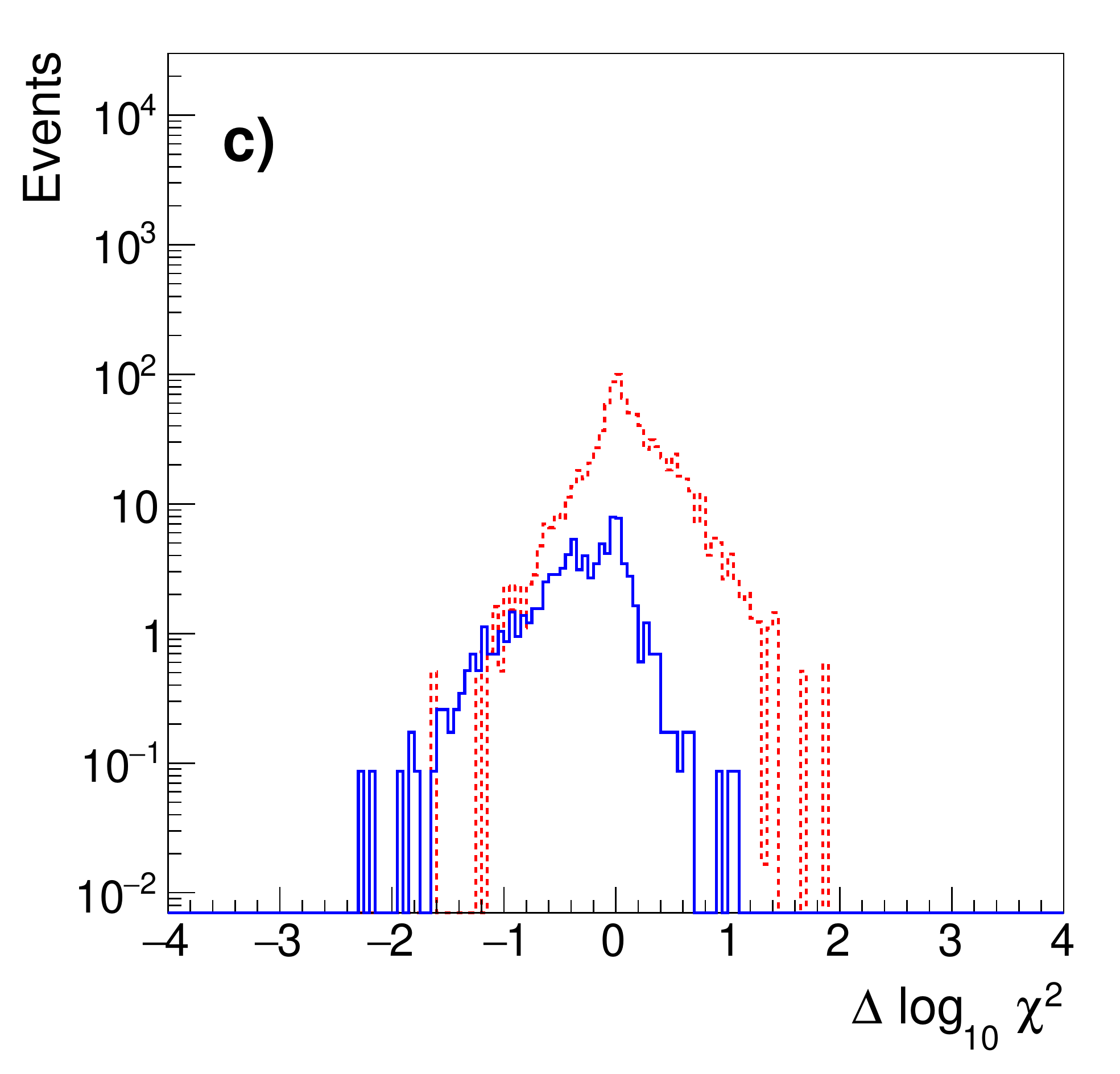}
\includegraphics[width=0.45\textwidth]{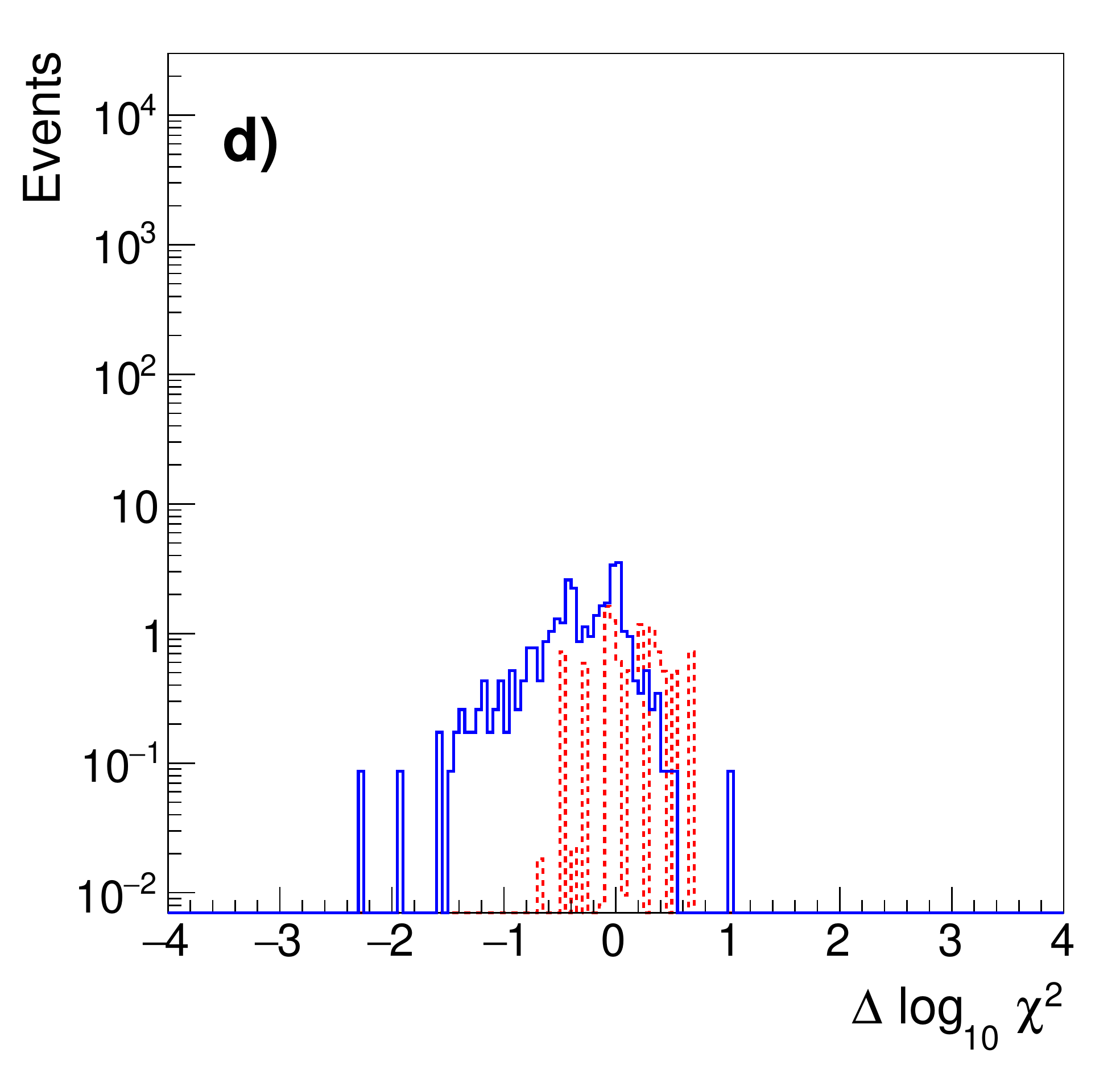}
\end{center}
\caption{\label{chi2_hists}
  Distributions of 
  $\Delta  \log_{10}\chi^2 = \log_{10} ( \chi^2_{\mathrm{sig}}/\chi^2_{\mathrm{bg}})$,
  for signal (solid blue histogram) and background (dashed red histogram)
  event samples, for all selected hadronic events (a), for
  events passing preselection cuts (b), for events with a good fit of
  the signal hypothesis (c) and the sample remaining after the final
  selection cuts (d). 
} 
\end{figure}

To discriminate between the background and the signal hypothesis the ratio
of the $\chi^2$ values of the two hypotheses was considered.
Shown in figure~\ref{chi2_hists} are the distributions of this ratio
(in logarithmic scale;
$\Delta  \log_{10}\chi^2 = \log_{10} ( \chi^2_{\mathrm{sig}}/\chi^2_{\mathrm{bg}})$),
for signal and background samples corresponding to the integrated
luminosity of 500~fb$^{-1}$.
Before any selection cuts (figure~\ref{chi2_hists}a), the background level
is three orders of magnitude higher than the signal one (which corresponds
to the assumed branching ratio for the FCNC decays).
Preselection cuts (figure~\ref{chi2_hists}b) and the
kinematic fit (figure~\ref{chi2_hists}c) allow for background
suppression by over two orders of magnitude, but it is still not
sufficient to identify the signal.
Also, the shape of the $\Delta \log_{10}\chi^2$ distribution is very
similar for signal and background samples, showing that the kinematic
fit itself is not sufficient for signal identification.
The kinematic reconstruction is clearly not precise enough.
Only after very tight cuts on the flavour tagging probabilities for
the three $b$-quark jets and $c$ jet, signal events can be selected
with a reasonable purity (figure~\ref{chi2_hists}d).
The final cut on  $\Delta \log_{10}\chi^2$ was then optimised to obtain
the best expected limit on the FCNC branching ratio.

For 500~fb$^{-1}$ of integrated luminosity and the optimised cut value
$\chi^2_{\mathrm{sig}}/\chi^2_{\mathrm{bg}} < 1.38$, the expected 95\% C.L. limit is

\[BR(t \rightarrow  c H ) \times BR(h  \rightarrow b \bar{b}) < 2.6
\cdot 10^{-4}\; .\]

The Standard Model background passing final selection cuts is
estimated to be about 4.9 events while 31.8 events are expected for
the signal (assuming $BR(t  \rightarrow  c H ) \times BR(h  
\rightarrow  b \bar{b}) = 10^{-3}$).
The corresponding selection efficiencies are 3.9\% for signal events
and $1.2\cdot 10^{-5}$ for the background sample.
Numbers of signal and background events expected, and corresponding selection
efficiencies after subsequent analysis cuts are summarised in table~\ref{seltab}.
All efficiencies are normalised to the total number of decays (including leptonic
and semi-leptonic decays).

\def\arraystretch{1.2}

\begin{table}
\begin{center}
  \begin{tabular}{|ll|r|r|r|r|}
    \hline
\multicolumn{2}{|l|}{Analysis level}  &  \multicolumn{2}{c|}{Expected events}  
                              &  \multicolumn{2}{c|}{Efficiency}     \\ \cline{3-6}
& Selection cut   & $t\bar{t}$ (SM) & Signal & $t\bar{t}$ (SM) & Signal  \\ \hline
\multicolumn{6}{|l|}{Input sample}  \\ \hline
 & Before any selection cut     & 410'000 & 819 & 100\% & 100\% \\ \hline
 \multicolumn{6}{|l|}{Preselection cuts (before kinematic fit)}  \\ \hline
 & $E_{\mathrm{balance}} < 100$ GeV &  167'000   & 499 & 40.6\% & 60.9\%  \\
 & 3 $b$ jets tagged {($b_{\mathrm{tag}}>0.4$)} & 13'280  & 300 & 3.24\% &  36.6\% \\
 & $c$ jet tagged {($b_{\mathrm{tag}} + c_{\mathrm{tag}} > 0.4$)} & 9640 & 276 & 2.35\% & 33.8\% \\ \hline
 \multicolumn{6}{|l|}{Final selection cuts (after kinematic fit)}  \\ \hline
& Good fit ($\chi^2_{\mathrm{sig}} < 14$, $\Delta M_{t} < 45$ GeV)
                      & 894 & 87 & 0.22\% & 10.7\% \\
& $b$-tag for higgs jets ($b_{1} \cdot b_{2}  > 0.95$)
          & 89.5 & 50.8 & 0.022\% & 6.2\%  \\
&  $b$ and $c$ tags
 ($b_{3}  > 0.9$, $c_{4}  + b_{4}  > 0.75$)
            & 10.7 & 34.1 & $2.6\! \cdot \!\! 10^{-5}$ & 4.2\%  \\
& $\chi^2_{\mathrm{sig}}/\chi^2_{\mathrm{bg}} < 1.38$  & 4.89 & 31.8 & $1.2\! \cdot\!\!  10^{-5}$  & 3.9\%  \\ \hline
\end{tabular}
\end{center}
\caption{\label{seltab}
  Numbers of signal and background events expected, and corresponding selection
  efficiencies after subsequent analysis cuts. Results based on the full detector
  simulation for $\sqrt{s} = 380$~GeV assuming an integrated luminosity of 500 fb$^{-1}$
  and $BR(t  \rightarrow  c H ) \times BR(h \rightarrow  b \bar{b}) = 10^{-3}$
  for the signal sample. Only the hadronic decay channel is considered in the analysis,
  while selection efficiencies are normalised to the total number of decays
  (including leptonic and semi-leptonic decays).}
\end{table}

\begin{figure}
\begin{center}
\includegraphics[width=0.6\textwidth]{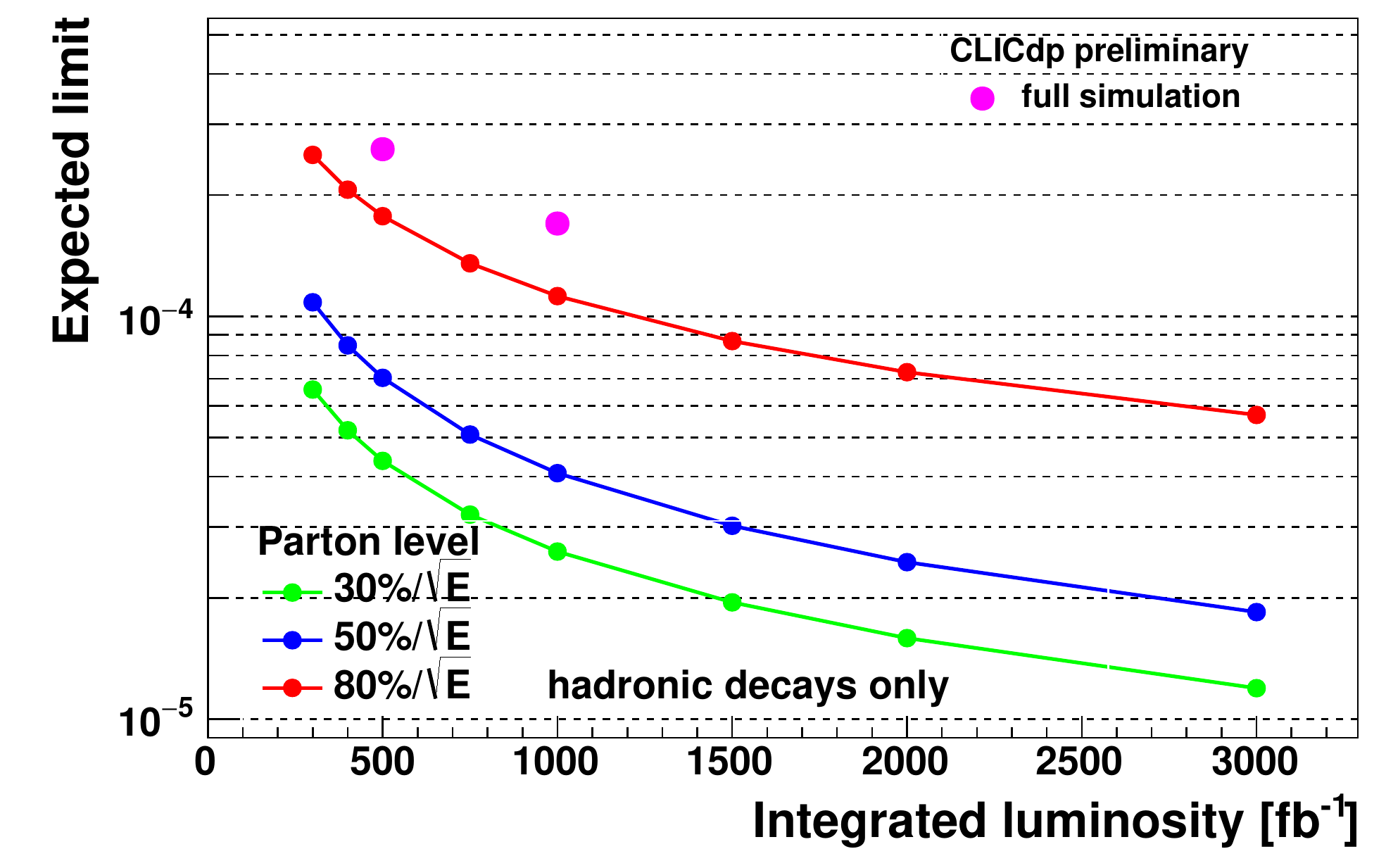}
\end{center}
\caption{\label{comp_limits}
  Expected 95\% C.L. limits on
  $BR(t  \rightarrow  c H ) \times BR(h \rightarrow  b  \bar{b})$
  as a function of the integrated luminosity for $e^+e^-$ collisions
  at $\sqrt{s} = 380$~GeV. Results based on the full detector
  simulation are compared to results from the parton level study
  restricted to the hadronic channel.} 
\end{figure}

\begin{figure}
\begin{center}
\includegraphics[width=0.6\textwidth]{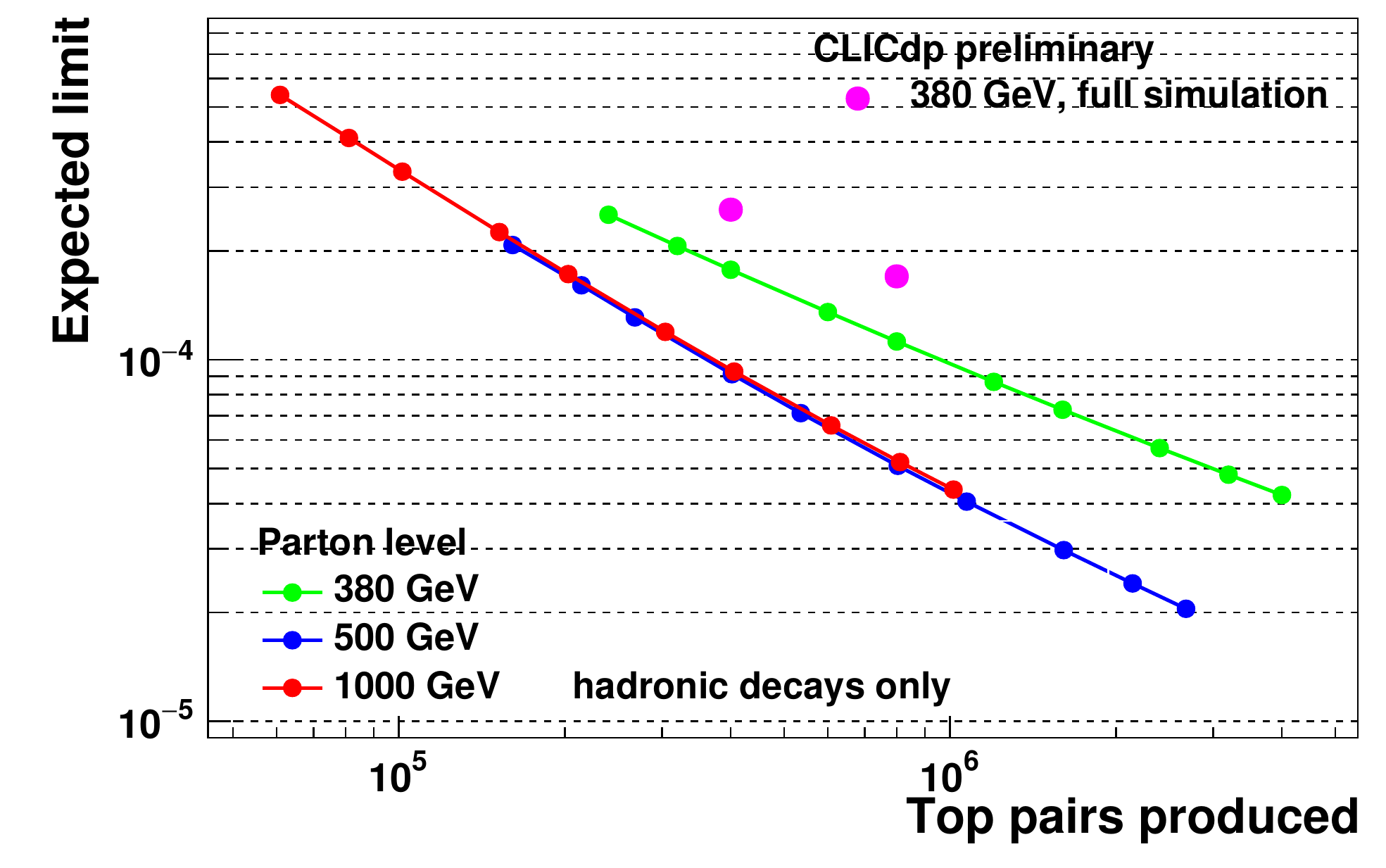}
\end{center}
\caption{\label{comp_limits2}
  Expected 95\% C.L. limits on
  $BR(t  \rightarrow  c H ) \times BR(h \rightarrow  b  \bar{b})$
  as a function of the collected sample of top pair production events.
  Results  based on the full detector simulation for $\sqrt{s} =
  380$~GeV are compared to  parton level results for different running
  energies, assuming jet energy resolution parameter $S = 80\%$.
  Only the hadronic decay channel is considered.} 
\end{figure}

Results from the detailed simulation study described in this
contribution are compared to the earlier estimates from the parton
level study in figures~\ref{comp_limits} and \ref{comp_limits2}.
For consistency, parton level results presented on these figures were obtained
considering the hadronic decay channel only.
The sensitivity of CLIC running at 380~GeV to FCNC top decays turns out to be
significantly weaker than the one estimated from the parton
level study, even assuming the worst (pessimistic) jet energy
resolution.
This is due to the much weaker discrimination power of the kinematic
fit.
Invariant mass resolutions for two- and three-jet configurations in
the final state ($W$ and top candiates), as obtained from the full
simulation study, are not well described by the Gaussian distributions.
Deterioration of the energy and mass resolution for the hadronic final
state can be due to the large fraction of $b$ jets in the considered
sample as well as to the influence of the overlaid beam background.
With the required background suppression factor of the order of
$10^{-4}$, tails of the mass distributions become very important and result
in a significant fraction of background events fitting signal hypothesis
better (two jets from $W$ decay looking like $H$ candidate).
These events can only be suppressed by very tight flavour tagging
cuts, but this reduces also the signal selection efficiency significantly.
A comparison of parton level results obtained for different running
energies, as shown in figure~\ref{comp_limits2}, indicates that the
performance of the kinematic fit should improve when going to higher
centre-of-mass energies.
At $\sqrt{s} = 380$~GeV, top quarks are produced almost at rest and
their decay products can easily mix.
At higher energies, a larger boost should result in better separation of
two top candidates and unambiguous result of the kinematic fit.

\section{Conclusions}

Considered in this contribution is  the feasibility of measuring the FCNC
top decay $t\rightarrow c H$ at CLIC running at 380 GeV.
Preliminary results based on the full detector simulation are presented.
The main focus of the analysis was on the optimization of the
kinematic event reconstruction in the hadronic channel.
After optimisation of the event selection criteria, the expected 95\%
C.L. limit on the FCNC branching ratio, assuming an integrated luminosity
of 500 fb$^{-1}$, is $2.6 \cdot 10^{-4}$.
This limit is significantly worse than the initial estimates based on
the parton level analysis, which is most likely due to the
non-Gaussian tails of the mass resolution distributions. 
The analysis is ongoing and different options will be studied to improve
signal to background discrimination and resulting limits.
This includes optimization of the LCFI+ performance, 
analysis of the semi-leptonic decay channel and  
possible use of MultiVariate Analysis tools to optimise event
selection.
Finally, measurements at higher collision energies should be also
considered.

\section*{Acknowledgments}

This work benefited from services provided by the ILC Virtual
Organisation, supported by the national resource providers of the EGI
Federation.



\printbibliography[title=References]

\end{document}